\documentclass[%
 reprint,
superscriptaddress,
 amsmath,amssymb,
 aps,
 prx, % PRX Quantum: REVTeX 4.2f has no prxquantum option, prx is the closest style
]{revtex4-2}

\usepackage{graphicx}% Include figure files
\usepackage{dcolumn}% Align table columns on decimal point
\usepackage{bm}% bold math
\usepackage{xcolor}
\usepackage{tikz}
\usepackage[noend]{algpseudocode}
\makeatletter
\newcounter{algorithm}
\renewcommand{\thealgorithm}{\arabic{algorithm}}
\def\fps@algorithm{tbp}
\def\ftype@algorithm{4}
\def\ext@algorithm{loa}
\def\fnum@algorithm{Algorithm~\thealgorithm}
\newenvironment{algorithm}{\@float{algorithm}}{\end@float}
\makeatother

\newcommand{\F}{\mathbb{F}_2}      % binary field
\newcommand{\xor}{\oplus}          % addition over F_2 of bits and bitvectors (XOR)
\newcommand{\band}{\land}          % bitwise AND
\newcommand{\Id}{\mathbb{I}}       % identity matrix (the identity operator is I)
\newcommand{\bv}[1]{\bm{#1}}       % bitvector, e.g. symplectic vector \bv{p} = (\xp{p} | \zp{p})
\newcommand{\xp}[1]{\bm{#1}^{x}}   % x-part of a symplectic vector
\newcommand{\zp}[1]{\bm{#1}^{z}}   % z-part of a symplectic vector
\newcommand{\xb}[2]{#1^{x}_{#2}}   % single x-bit on site #2
\newcommand{\zb}[2]{#1^{z}_{#2}}   % single z-bit on site #2
\newcommand{\partition}[1]{\mathcal{B}_{#1}} % partition with label #1
\newcommand{\ket}[1]{\lvert #1 \rangle}

\DeclareMathOperator{\rank}{rank}
\newcommand{\hw}[1]{\lvert #1 \rvert} % Hamming weight of a bitvector
\DeclareMathOperator{\im}{im}
\DeclareMathOperator{\Span}{span}
\usetikzlibrary{arrows.meta, calc}

\providecommand{\F}{\mathbb{F}_2}
\providecommand{\xor}{\oplus}
\providecommand{\bv}[1]{\bm{#1}}
\providecommand{\partition}[1]{\mathcal{B}_{#1}}

\definecolor{ppbit}{HTML}{F7D58B}   % set bit of a Pauli string
\definecolor{pphash}{HTML}{EC705A}  % set bit of the hash matrix
\definecolor{pplabel}{HTML}{9EC6A9} % updated label
\definecolor{ppblue}{HTML}{D2D6FB}  % highlighted partition
\definecolor{ppgrey}{HTML}{E9E9E9}  % partition
\definecolor{ppdelta}{HTML}{D0281C} % label shift Mq, multiplied string
\definecolor{ppkeep}{HTML}{FA8072}  % unchanged string

\newlength{\ppcell}
\tikzset{
    pp/picture/.style={x=\ppcell, y=\ppcell, font=\footnotesize},
    pp/cell/.style={line width=0.6pt},
    pp/box/.style={draw, line width=0.6pt, rounded corners=2.5pt},
    pp/arrow/.style={-{Stealth[length=3.5pt, width=3pt]}, line width=0.6pt},
}

\newcommand{\ppbits}[3][ppbit]{%
    \begin{scope}[shift={#2}]
        \edef\pplist{#3}
        \foreach \b [count=\i from 0] in \pplist {
            \ifnum\b=1 \fill[#1] (\i,0) rectangle ++(1,1); \else \fill[white] (\i,0) rectangle ++(1,1); \fi
            \draw[pp/cell] (\i,0) rectangle ++(1,1);
        }
    \end{scope}}

\newcommand{\ppbitscol}[3][ppbit]{%
    \begin{scope}[shift={#2}]
        \edef\pplist{#3}
        \foreach \b [count=\i from 1] in \pplist {
            \ifnum\b=1 \fill[#1] (0,-\i) rectangle ++(1,1); \else \fill[white] (0,-\i) rectangle ++(1,1); \fi
            \draw[pp/cell] (0,-\i) rectangle ++(1,1);
        }
    \end{scope}}

\def\ppboxh{3.8}
\def\ppboxgap{0.8}
\def\ppboxl{-3.9}
\def\ppboxr{10.1}

\newcommand{\ppbox}[4]{%
    \pgfmathsetmacro{\pptop}{-#2*(\ppboxh+\ppboxgap)}
    \pgfmathsetmacro{\ppx}{0.5*(\ppboxl+\ppboxr)-3}

    \draw[pp/box, fill=#1] (\ppboxl,\pptop) rectangle (\ppboxr,\pptop-\ppboxh);
    \coordinate (part-#3-west) at (\ppboxl,\pptop-0.5*\ppboxh);
    \coordinate (part-#3-east) at (\ppboxr,\pptop-0.5*\ppboxh);

    \foreach \ppc/\ppb/\pps [count=\j from 1] in {#4} {
        \pgfmathsetmacro{\ppy}{\pptop-0.25-1.5*\j}
        \xdef\pplast{\ppy}
        \ppbits{(\ppx,\ppy)}{\ppb}
        \node[anchor=east,inner sep=1.5pt] at (\ppx,\ppy+0.5) {$c_{\ppc}$};
        \node[anchor=west,inner sep=1.5pt] at (\ppx+6,\ppy+0.5) {$\pps$};
    }

    \draw[pp/cell] (\ppx+3,\pptop-0.45) -- (\ppx+3,\pplast-0.3);
}
\newcommand{\ppstack}[4]{%
    \ppbox{#1}{0}{00}{{00,1}/{1,0,0,1,0,1}/{Y_1Z_3}}
    \ppbox{#2}{1}{01}{{01,1}/{0,0,0,1,0,0}/{Z_1}}
    \ppbox{#3}{2}{10}{{10,1}/{0,0,0,0,1,0}/{Z_2}, {10,2}/{1,1,0,1,0,0}/{Y_1X_2}}
    \ppbox{#4}{3}{11}{{11,1}/{0,0,1,0,0,0}/{X_3}}}
\usepackage{stackengine}

\newcommand{\secref}[1]{Sec.~\ref{#1}}
\renewcommand{\eqref}[1]{Eq.~(\ref{#1})} % overrides the amsmath version
\newcommand{\tableref}[1]{Table~\ref{#1}}
\newcommand{\figref}[1]{Fig.~\ref{#1}}
\newcommand{\appref}[1]{App.~\ref{#1}}
\newcommand{\panellabel}[4]{%
  \stackinset{l}{#2}{t}{#3}{\small(#1)}{#4}%
}

\begin{document}

\title{Scalable Pauli propagation through $\F$-linear partitioning}% Force line breaks with \\

\author{Alev Orfi}
    \affiliation{Center for Computational Quantum Physics, Flatiron Institute, 162 5th Avenue, New York, NY 10010, USA}
    \affiliation{Department of Physics, New York University, 726 Broadway, New York, NY 10003, USA}
\author{Andrew Kille}
    \affiliation{Center for Computational Quantum Physics, Flatiron Institute, 162 5th Avenue, New York, NY 10010, USA}
    \affiliation{Department of Physics, New York University, 726 Broadway, New York, NY 10003, USA}
\author{J. Clayton Peacock}
    \affiliation{Department of Physics, Columbia University, New York, NY 10027, USA}
\author{Jack Dunham}
    \affiliation{Center for Computational Quantum Physics, Flatiron Institute, 162 5th Avenue, New York, NY 10010, USA}
\author{Lukas Devos}
    \email{ldevos98@gmail.com}
    \affiliation{Center for Computational Quantum Physics, Flatiron Institute, 162 5th Avenue, New York, NY 10010, USA}
\date{\today}

\begin{abstract}
Pauli propagation simulates quantum circuits and many-body dynamics by evolving observables in the Heisenberg picture as sparse sums of Pauli strings.
As Pauli-string operations reduce to bit manipulations, its performance is limited by memory bandwidth and capacity rather than computational throughput.
Pooling the memory of many cores and nodes is hindered by the deduplication step after every gate, which can require global synchronization.
We avoid this bottleneck by partitioning the strings with an $\F$-linear hash function, which guarantees that a $k$-qubit channel couples each partition to at most $4^k$ partitions fixed by the gate, decomposing gate application into independent tasks with bounded communication.
Benchmarks of our open-source implementation demonstrate efficient parallel scaling within and across compute nodes, enabling simulations with over $10^{11}$ Pauli strings and extending the accessible timescales of quantum many-body dynamics.
\end{abstract}

\maketitle

%\tableofcontents
\section{Introduction}
%edit in reviewing mode

Classical simulation methods are central to the study of quantum many-body systems and the development of quantum technologies, providing both theoretical insight and benchmarks for quantum experiments.
For example, the kicked Ising experiment on a $127$-qubit superconducting processor~\cite{kim2023evidence} was quickly followed by classical simulations that matched or surpassed its accuracy~\cite{tindall2024efficient, begusic2024fast,liao2023simulation,anand2023classical,patra2024efficient}.
Among these methods is \emph{Pauli propagation}, which evolves observables in the Heisenberg picture as sums of Pauli strings.
Its cost depends only on the number of terms, which Clifford gates leave unchanged, making it particularly effective for near-Clifford and certain noisy circuits, where its efficiency is rigorously established~\cite{aharonov2023polynomial, schuster2025polynomial, fontana2025classical, angrisani2025classically, angrisani2026simulating}, and it has also proven effective for the real-time dynamics of local observables~\cite{begusic2025realtime, loizeau2025quantum}.

Pauli propagation can be implemented efficiently using the \emph{binary symplectic representation}, which encodes each Pauli string as a pair of bitstrings and reduces multiplication and commutation checks to simple bitwise operations~\cite{Dehaene2003,rall2019simulation,gidney2021stim,loizeau2025quantum, rudolph2025pauli, krotz2026paulib,broers2026exclusive,broers2026scalable}.
Since individual string operations are cheap, large-scale simulations are limited by memory: its bandwidth sets the propagation rate~\cite{wulf1995hitting}, and its capacity the number of strings that can be retained, and hence the achievable accuracy.
This motivates distributing the expansion over many processors and nodes~\cite{broers2026scalable}.
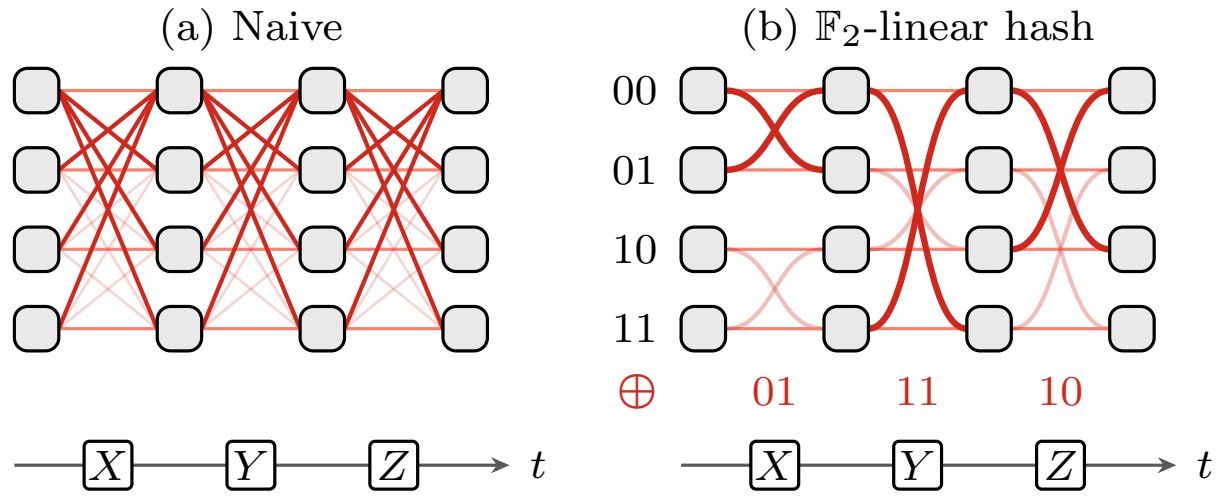
\begin{figure}[h]
    \centering
    % Naive versus F_2-linear partitioning with two label bits over three gates, see fig:idea in main.tex.
\begin{tikzpicture}[pp/picture]
        % (a) naive split at x = 0, (b) F_2-linear hash at x = 21; columns are gates, rows partitions
    \node[anchor=base] at (6.75,1.6) {(a) Naive};
    \node[anchor=base] at (27.75,1.6) {(b) $\F$-linear hash};
    \foreach \l [count=\r from 0] in {00,01,10,11}
        \node[anchor=east] at (20.1,-2.5*\r) {$\l$};

    % label shift of each gate, below the gate, with xor as the row label
    \node[anchor=east, text=ppdelta] at (20.1,-9.47) {$\bm{\xor}$};
    \foreach \l [count=\s from 0] in {01,11,10}
        \node[anchor=base, text=ppdelta] at (21+4.5*\s+2.25,-9.9) {$\l$};

    % time runs from left to right; the gates are rotations generated by X, Y and Z
    \foreach \p in {0,21} {
        \draw[pp/arrow, black!65] (\p-0.7,-11.8) -- (\p+14.9,-11.8);
        \foreach \g [count=\s from 0] in {X,Y,Z}
            \node[draw, fill=white, pp/cell, rounded corners=1pt, minimum size=9pt, inner sep=0pt]
                at (\p+4.5*\s+2.25,-11.8) {$\g$};
        \node[anchor=west] at (\p+15,-11.8) {$t$};
    }
    \path (-2.9,0);
    
    % (a) every partition sends products to every other partition (faded)
    \foreach \s in {0,1,2}
        \foreach \r in {0,...,3} {
            \foreach \t in {0,...,3} {
                \ifnum\t=\r\else
                    \draw[ppdelta, line width=0.5pt, opacity=0.2] (4.5*\s+0.7,-2.5*\r) -- (4.5*\s+3.8,-2.5*\t);
                \fi
            }
            \draw[pp/cell, ppkeep] (4.5*\s+0.7,-2.5*\r) -- (4.5*\s+3.8,-2.5*\r);
        }

    % (b) every partition exchanges with the partner whose label differs by the gate's shift (faded),
    % partner = r xor d on two bits
    \foreach \s/\d in {0/1, 1/3, 2/2}
        \foreach \r in {0,...,3} {
            \pgfmathtruncatemacro{\partner}{mod(\r+\d,2)+2*mod(floor(\r/2)+floor(\d/2),2)}
            \draw[pp/cell, ppkeep] (21+4.5*\s+0.7,-2.5*\r) -- (21+4.5*\s+3.8,-2.5*\r);
            \draw[ppdelta, line width=0.8pt, opacity=0.3]
                (21+4.5*\s+0.7,-2.5*\r) .. controls +(1.55,0) and +(-1.55,0) .. (21+4.5*\s+3.8,-2.5*\partner);
        }

    % partition 00 highlighted, both directions: every other partition in (a), one partner in (b)
    \foreach \s in {0,1,2}
        \foreach \t in {1,2,3} {
            \draw[ppdelta, line width=0.8pt] (4.5*\s+0.7,0) -- (4.5*\s+3.8,-2.5*\t);
            \draw[ppdelta, line width=0.8pt] (4.5*\s+0.7,-2.5*\t) -- (4.5*\s+3.8,0);
        }
    \foreach \s/\d in {0/1, 1/3, 2/2} {
        \draw[ppdelta, line width=1.2pt]
            (21+4.5*\s+0.7,0) .. controls +(1.55,0) and +(-1.55,0) .. (21+4.5*\s+3.8,-2.5*\d);
        \draw[ppdelta, line width=1.2pt]
            (21+4.5*\s+0.7,-2.5*\d) .. controls +(1.55,0) and +(-1.55,0) .. (21+4.5*\s+3.8,0);
    }

    % partitions
    \foreach \p in {0,21}
        \foreach \s in {0,...,3}
            \foreach \r in {0,...,3}
                \draw[pp/box, fill=ppgrey] (\p+4.5*\s-0.7,-2.5*\r+0.7) rectangle ++(1.4,-1.4);
\end{tikzpicture}
    \caption{Naive versus $\F$-linear partitioning for successive Pauli rotations.
    Grey squares denote partitions between gates, and red lines indicate possible exchanges.
    (a) Naive partitioning permits near-all-to-all communication.
    (b) $\F$-linear hashing shifts partition labels by a gate-dependent XOR offset, restricting exchanges to predetermined pairs.}
    \label{fig:overview}
\end{figure}

Although strings can be evolved independently, different inputs can produce identical outputs, whose coefficients must be summed in a \emph{deduplication step}.
In a naive parallel implementation, this requires global communication and synchronization, introducing an overhead that grows with the number of processes.
To avoid this, we partition the strings with an $\F$-linear hash function.
By linearity, multiplication by a fixed Pauli string maps every partition as a whole onto a single other partition.
Since a Pauli rotation combines this multiplication with the identity, each partition couples to at most one other, replacing global communication with independent pairwise exchanges and allowing deduplication to be performed locally, as illustrated in \figref{fig:overview}.
The number of partitions can be chosen independently of the number of workers, allowing task sizes to be optimized for cache locality and improving performance even on a single core.

We benchmark our open-source implementation, \texttt{paulistrings-rs}~\footnote{\texttt{https://github.com/lkdvos/paulistrings-rs}}, on the kicked Ising circuit of Ref.~\cite{kim2023evidence} and quenches of the two-dimensional transverse-field Ising model~\cite{begusic2025realtime,vovrosh2026transverseising}.
Within a node, the implementation exhibits near-ideal scaling until memory bandwidth becomes limiting.
Across nodes, it achieves substantial parallel speedups while maintaining weak-scaling efficiencies of $68$--$74\%$ at $32$ nodes, supporting aggregate budgets of up to $1.6\times10^{11}$ Pauli strings.
By exploiting the combined memory capacity of multiple nodes, the implementation enables larger Pauli expansions and extends the accessible timescales of classical quantum dynamics simulations.

\section{Methods}
\label{sec:methods}

Pauli propagation represents operators as sparse sums of Pauli strings, 
\begin{equation}\label{eq:PauliExpand}
    O = \sum_{i=1}^N c_i P_i,
\end{equation}
where every $P_i \in \{I,X,Y,Z\}^{\otimes L}$ for a system of $L$ qubits.
Pauli strings span the $4^L$-dimensional operator space, and many operators admit sparse representations with $N\ll4^L$. 
Pauli propagation evolves these expansions through successive gate applications, updating the strings and their coefficients. 
This approach remains efficient when the number of strings required to approximate the operator of interest remains manageable, as is often the case for a variety of applications~\cite{loizeau2025quantum,begusic2025realtime,rudolph2025pauli}. 

To manipulate these expansions efficiently, we use the \emph{symplectic encoding} of Pauli strings~\cite{Dehaene2003}.
Each single-qubit Pauli operator is represented by two bits $(x,z)\in\F^{~2}$, with the phase convention $P(x,z)=i^{xz}X^xZ^z$ ensuring Hermiticity.
For $L$ qubits, these bits form a vector over the binary field, $\bv{p}=(\xp{p}\mid\zp{p})\in\F^{~2L}$, representing the string
\begin{equation}
    P(\bv{p}) = i^{N_Y(\bv{p})}
    \bigotimes_{j=1}^L X^{\xb{p}{j}}Z^{\zb{p}{j}},
\end{equation}
where $N_Y(\bv{p})=\hw{\xp{p}\band\zp{p}}$ counts the number of $Y$ factors, $\band$ denotes bitwise AND, and $\hw{\cdot}$ the Hamming weight.

In this encoding, multiplication of two Pauli strings reduces to bitwise XOR $\xor$, up to a phase,
\begin{equation}
    \label{eq:stringmult}
    P(\bv{p})P(\bv{q})
    = i^{\kappa(\bv{p},\bv{q})}P(\bv{p}\xor\bv{q}),
\end{equation}
where the phase follows from commuting $Z^{\zp{p}}$ past $X^{\xp{q}}$, which gives a minus sign wherever both act,
\begin{equation}
    \kappa(\bv{p},\bv{q}) = N_Y(\bv{p}) + N_Y(\bv{q}) - N_Y(\bv{p}\xor\bv{q}) + 2\hw{\zp{p}\band\xp{q}}.
\end{equation}
Commutation is determined by the symplectic product
\begin{equation}
    \label{eq:symprod}
    s(\bv{p},\bv{q}) = \bv{p}^TJ\bv{q},
    \qquad
    J=\begin{pmatrix}0&\Id\\ \Id&0\end{pmatrix},
\end{equation}
such that $PQ=(-1)^{s(\bv{p},\bv{q})}QP$, where the symplectic form $J$ over $\F$ gives the encoding its name.

A gate generally maps each input Pauli string to a linear combination of output strings, with the number of terms defining the \emph{branching ratio}.
A Pauli rotation $R_Q(\theta) = \exp(-i\frac{\theta}{2}Q)$ transforms a string $P$ depending on whether it commutes or anticommutes with $Q$:
\begin{equation}
\label{eq:paulirotation}
    R_Q^\dagger(\theta) P R_Q(\theta) =
        \begin{cases}
            P, & s(\bv{p}, \bv{q}) = 0,\\
            \cos\theta\,P + i\sin\theta\,QP, & s(\bv{p}, \bv{q}) = 1.
        \end{cases}
\end{equation}
Pauli rotations, with a branching ratio of at most $2$, are the elementary gates used in the Trotterized simulations considered below.

After each gate application, the output strings must be collected into an expansion of the form of \eqref{eq:PauliExpand}, with each string appearing once.
Duplicate Pauli strings can be generated by the gate, and these must then be identified and their coefficients summed in a \emph{deduplication step}. 
When gate applications are parallelized, strings are distributed across processes, but naive partitioning can generate duplicate strings on different processes, in which case deduplication requires global communication.
To avoid this, we introduce an $\F$-linear hashing scheme that decomposes gate applications into independent tasks involving a bounded number of partitions, enabling predictable communication and local deduplication.

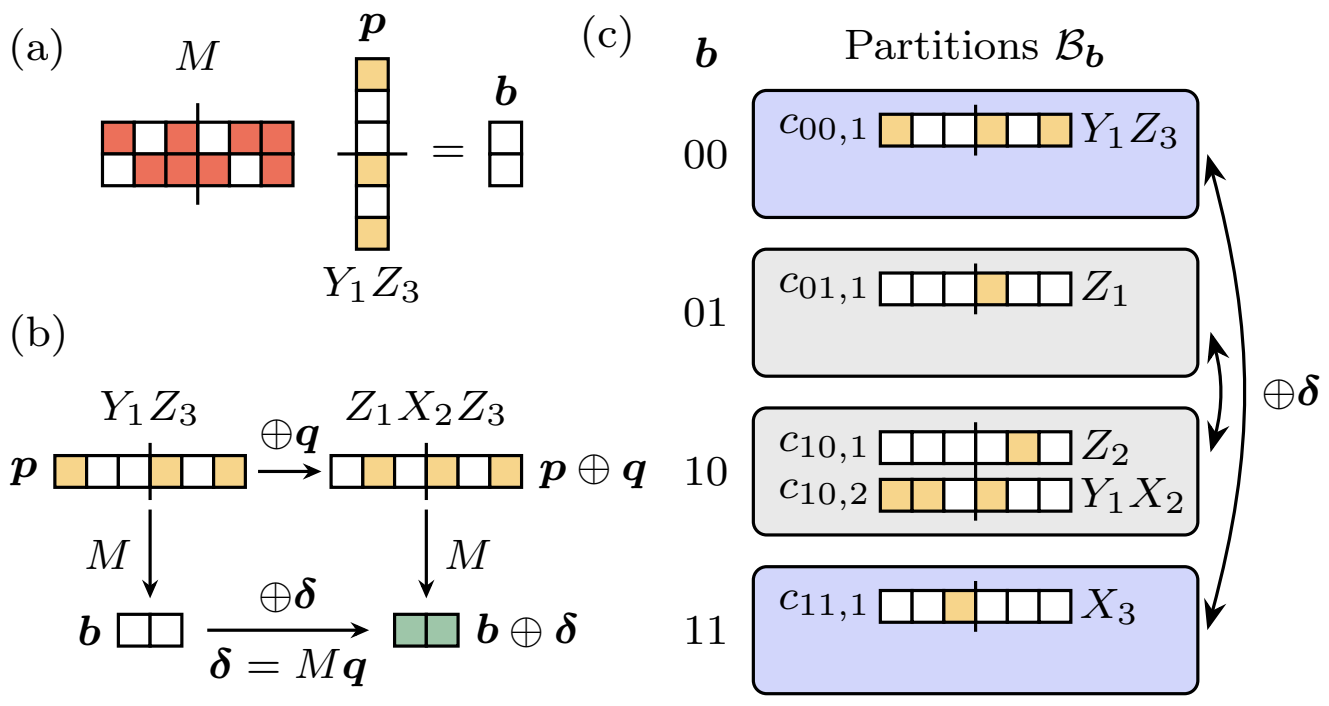
\begin{figure}[t]
    \centering
    \begin{tikzpicture}[pp/picture]
    \node[anchor=north west,inner sep=0pt] at (-2,1) {(a)};
    \node[anchor=north west,inner sep=0pt] at (-2,-8) {(b)};
    % \node[anchor=north west,inner sep=0pt] at (-2,-20) {(c)};
    \node[anchor=north west,inner sep=0pt] at (16,1.4) {(c)};

    % (a) Hash assignment
    \begin{scope}[shift={(1,5)}]
        \ppbits[pphash]{(0,-8)}{1,0,1,0,1,1}
        \ppbits[pphash]{(0,-9)}{0,1,1,1,0,1}
        \draw[pp/cell] (3,-9.6)--(3,-6.4);
        \node[anchor=south,yshift=3pt] at (3,-6.4) {$M$};

        \ppbitscol{(8,-5)}{1,0,0,1,0,1}
        \draw[pp/cell] (7.4,-8)--(9.6,-8);
        \node[anchor=south] at (8.5,-5) {$\bv{p}$};
        \node[anchor=north] at (8.5,-11) {$Y_1Z_3$};

        \node at (10.9,-8) {$=$};
        \ppbitscol[white]{(12.2,-7)}{0,0}
        \node[anchor=south] at (12.7,-7) {$\bv{b}$};
    \end{scope}

    % % (b) Pauli string q
    % \begin{scope}[shift={(0,-9)}]
    %     \ppbits{(4.5,-1.5)}{1,1,0,0,0,0}
    %     \draw[pp/cell] (7.5,-1.9)--(7.5,0.1);
    %     \node[anchor=east] at (4.5,-1) {$\bv{q}$};
    %     \node[anchor=west] at (10.5,-1) {$X_1X_2$};
    % \end{scope}
    
    % (c) Multiplication and hash translation
    \begin{scope}[shift={(-0.5,-6)}]
        \ppbits{(0,-7.5)}{1,0,0,1,0,1}
        \ppbits{(8.7,-7.5)}{0,1,0,1,0,1}
        \draw[pp/cell] (3,-7.9)--(3,-6.1) (11.7,-7.9)--(11.7,-6.1);
        \node[anchor=south] at (3,-6.1) {$Y_1Z_3$};
        \node[anchor=south] at (11.7,-6.1) {$Z_1X_2Z_3$};
        \node[anchor=east] at (0,-7) {$\bv{p}$};
        \node[anchor=west] at (14.7,-7) {$\bv{p}\xor\bv{q}$};
        \draw[pp/arrow] (6.4,-7)--node[above] {$\xor\bv{q}$} (8.4,-7);
    
        \draw[pp/arrow] (3,-8.2)--node[left] {$M$} (3,-11);
        \draw[pp/arrow] (11.7,-8.2)--node[right] {$M$} (11.7,-11);
        \ppbits[white]{(2,-12.5)}{0,0}
        \ppbits[pplabel]{(10.7,-12.5)}{1,1}
        \node[anchor=east] at (2,-12) {$\bv{b}$};
        \node[anchor=west] at (12.7,-12) {$\bv{b}\xor\bv{\delta}$};
        \draw[pp/arrow] (4.8,-12)--node[above] {$\xor\bv{\delta}$}
            node[below] {$\bv{\delta}=M\bv{q}$} (9.9,-12);
    \end{scope}

    % (d) Partition exchanges
    \begin{scope}[shift={(24,-1)}]
        \def\ppboxh{4}
        \def\ppboxgap{1}
        \def\ppboxl{-2.5}
        \def\ppboxr{11.5}
    
        \ppstack{ppblue}{ppgrey}{ppgrey}{ppblue}
        \node[anchor=south] at (0.5*\ppboxl+0.5*\ppboxr,0.2) {Partitions $\partition{\bv{b}}$};
    
        \node[anchor=south] at (\ppboxl-1.5,0.2) {$\bv{b}$};
        \foreach \l [count=\k from 0] in {00,01,10,11}
            \node at (\ppboxl-1.5,-\k*\ppboxh-\k*\ppboxgap-.5*\ppboxh) {$\l$};
    
        \draw[pp/arrow,Stealth-Stealth]
            ($(part-00-east)+(.3,0)$) to[bend left=15]
            node[right] {$\xor\bv{\delta}$}
            ($(part-11-east)+(.3,0)$);
    
        \draw[pp/arrow,Stealth-Stealth]
            ($(part-01-east)+(.4,-.6)$) to[bend left=15]
            ($(part-10-east)+(.4,.6)$);
    \end{scope}
\end{tikzpicture}
    \caption{ $\F$-linear partitioning and multiplication.
    (a) The Pauli string $P = Y_1 Z_3$ has bit vector $\bv{p} = (\xp{p}\mid\zp{p}) \in \F^{2L}$.
    The binary hash matrix $M$ assigns it the label $\bv{b} = M\bv{p}$, computed mod $2$, and hence the partition $\partition{\bv{b}}$.
    Each partition stores the strings $\bv{p}_{\bv{b},i}$ and coefficients $c_{\bv{b},i}$ of the operator $O_{\bv{b}}$.
    (b)--(c) Multiplication by $Q = X_1X_2$, with bitvector $\bv{q}$, maps $\bv{p}$ to $\bv{p} \xor \bv{q}$ and the label $\bv{b}$ to $\bv{b} \xor \bv{\delta}$ with $\bv{\delta} = M\bv{q}$ (b).
    Every partition is therefore mapped onto a single partner, and the partitions are exchanged in pairs (c).
    Shown here is the case $L = 3$, $B = 2$.
    Labels are written as $\bv{b} = b_1 b_2$, and filled (empty) squares denote $1$ ($0$).}
    \label{fig:hash}
\end{figure}

In this scheme, as shown in \figref{fig:hash}(a), a binary matrix $M\in\F^{B\times 2L}$ assigns each Pauli string to one of $2^B$ partitions,
\begin{equation}
    \partition{\bv{b}} = \left\{ P(\bv{p}) ~\vert~ M\bv{p} = \bv{b} \right\}, \quad \bv{b} \in \F^B,
\end{equation}
and by linearity $M(\bv{p}\xor\bv{q})= M\bv{p}\xor M\bv{q}$.
Consequently, multiplication by a fixed Pauli string $Q$ maps every string in $\partition{\bv{b}}$ to $\partition{\bv{b}\xor M\bv{q}}$, as illustrated in \figref{fig:hash}(b).
Unlike the hash-based distribution in ORQA~\cite{broers2026scalable}, this restricts communication to a single predetermined partner per partition, avoiding string-dependent routing, as shown in \figref{fig:hash}(c).

A Pauli rotation $R_Q(\theta)$ maps each string $P\in\partition{\bv{b}}$ to a combination of $P$ and $QP$ as shown in \eqref{eq:paulirotation}.
The former remains in $\partition{\bv{b}}$, while the latter belongs to $\partition{\bv{b}\xor M\bv{q}}$.
Therefore, each partition interacts with at most one other partition, forming independent pairs that can be updated and deduplicated separately.

More generally, any gate or channel acting on $k$ qubits maps $P(\bv{p})$ to a linear combination of strings $P(\bv{p}\xor\bv{d})$, where $\bv{d}$ is supported only on those qubits.
By linearity of the hash, the output partition label is $M(\bv{p}\xor\bv{d})=\bv{b}\xor M\bv{d}$.
Let $W_U\subseteq \F^{2L}$ denote the span of all such offsets, and $D_U = MW_U$ their image under the hash.
Since the offsets involve at most $2k$ binary coordinates, 
\begin{equation}
    d_U = \dim D_U \leq \min(B, 2k).
    \label{eq:offsetdimension}
\end{equation}
Each partition $\partition{\bv{b}}$ can therefore exchange strings only with partitions labeled by $\bv{b}\xor D_U$.
These labels form a coset of $D_U$, which is closed under all offsets generated by the gate.
Consequently, the $2^B$ partitions split into $2^{B-d_U}$ disjoint groups, each containing $2^{d_U}\leq4^k$ partitions.
Each group forms an independent task that can be updated and deduplicated without communication with other groups.

The choice of $M$ involves a tradeoff between balancing the number of strings across partitions and minimizing communication between them. 
A uniformly random full-rank matrix provides approximately universal hashing, with distinct strings colliding with probability close to $2^{-B}$~\cite{carter1979universal}, yielding approximately balanced partitions when $N\gg2^B$. 
However, $M$ can also be optimized for the specific simulation and hardware, for example to reduce the communication between compute nodes.
We discuss this trade-off in more detail in \appref{app:hash}.

The decomposition into independent groups of work provides the basis for both shared-memory parallelism as well as distributed-memory communication.
Within a node, these tasks are distributed over the threads by work stealing scheduling, which balances their uneven sizes.
Since tasks update disjoint partitions, no locking is required, and threads synchronize only after completing each gate.
For distributed execution, additional hash bits assign partitions to NUMA domains or MPI processes, $\bv{b}=(\bv{b}_{\mathrm{node}}\mid\bv{b}_{\mathrm{local}})$.
The gate-dependent offsets determine the communication partners in advance, enabling bounded point-to-point exchanges.
Truncation and measurement require at most a scalar or histogram reduction over the partitions, as discussed in \appref{app:parallel_ops}.
This partitioning also benefits single-worker performance by reducing the deduplication working set to fit within processor caches, as discussed in \appref{app:engine}.
Together, these features enable effective use of available computing resources, as demonstrated in \secref{sec:results}.

\section{Results}
\label{sec:results}

\subsection{Kicked Ising dynamics}
\label{sec:kickedising}

\begin{figure}[t]
{\includegraphics[width=\linewidth]{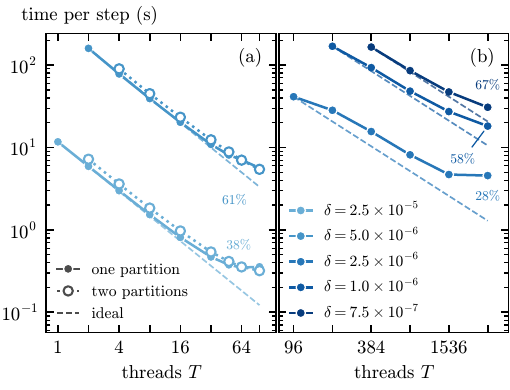}}
\caption{Strong scaling of the wall time of $\langle Z_{62}\rangle$ for the kicked Ising circuit ($127$ qubits, $\theta_h=\pi/4$, $20$ Trotter steps).
Dashed lines indicate ideal scaling, with percentages giving the corresponding parallel efficiencies relative to the first point of each curve. 
(a) Scaling on a single two-socket AMD EPYC 9474F node, for a single pool of threads (one partition) and with two partitions assigned to the two NUMA domains by a lattice-cut node hash.
(b) Scaling across multiple nodes, with one MPI rank per NUMA domain partitioned according to the avoid $50\%$ hash (see \appref{app:noderows}).
Each curve begins at the smallest number of nodes whose combined memory holds the operator.}
\label{fig:single_multi_node}
\end{figure}

We first use our implementation to apply a kicked Ising circuit on the $127$-qubit heavy-hex lattice introduced in Ref.~\cite{kim2023evidence}, a standard benchmark for Pauli propagation methods~\cite{begusic2024fast,rudolph2025pauli,mazumder2026benchmarking}.
Every Trotter step applies the unitary circuit
\begin{equation}
    U = \left[\prod_{\langle jk\rangle} \exp\left(i\frac{\pi}{4} Z_j Z_k\right)\right] \left[\prod_j \exp\left(-i\frac{\theta_h}{2} X_j\right)\right],
\end{equation}
where $\langle jk\rangle$ denotes nearest-neighbor pairs, and $\theta_h$ is the angle of the transverse-field kick, here fixed to $\theta_h=\pi/4$. 
Using the initial product state $\ket{0}^{\otimes 127}$, we calculate the single-site magnetization $\langle Z_{62} \rangle$ after $20$ Trotter steps. 
Coefficients are truncated using an absolute threshold $|c_i|
\leq \delta$, with alternative truncation methods detailed in \appref{app:parallel_ops}.

We first examine the strong scaling on a single two-socket node, where ideal performance corresponds to wall time inversely proportional to the number of threads. \figref{fig:single_multi_node}(a) shows the wall time per Trotter step as a function of thread count, for $\delta = 2.5\times 10^{-5}$ and $\delta = 5.0\times 10^{-6}$.
The single-partition configuration uses a shared thread pool, while the two-partition configuration assigns the expansion to separate NUMA domains using the lattice-cut hash of \appref{app:noderows}.
The two-partition configuration is used for the multi-node benchmarks below.
Efficiencies are measured relative to the first point of the curve, at $T = 1$ and $T = 2$, respectively.

Scaling is close to ideal up to $16$ threads, after which efficiency gradually decreases, reaching $61\%$ for $\delta = 5\times 10^{-6}$ and $38\%$ for $\delta = 2.5\times 10^{-5}$ at $96$ threads.
The deviation from ideal is consistent with saturation of the available memory bandwidth, as discussed in \appref{app:engine}.
The larger cutoff produces a smaller Pauli expansion, so fixed per-gate and per-task overheads account for a larger fraction of its wall time, causing an earlier deviation from ideal scaling.
The two-partition configuration is slower at low thread counts, where the single-partition configuration benefits from keeping the entire expansion within one NUMA domain, but becomes faster at $64$ threads and above by improving memory locality. 

We next examine strong scaling across multiple nodes, where aggregate memory makes smaller thresholds accessible.
\figref{fig:single_multi_node}(b) shows the wall time as a function of node count for several $\delta$ using two MPI ranks per node (one per NUMA domain) and the \emph{avoid} $f = 50\%$ node hash of \appref{app:noderows}.
Each curve starts at the minimum number of nodes required to accommodate the operator, increasing from one node at $\delta = 2.5\times 10^{-6}$ to four nodes for $\delta = 7.5\times 10^{-7}$. 
For the three thresholds, wall time decreases monotonically, reaching efficiencies of $28\%$, $58\%$, and $67\%$, respectively.
At a fixed number of nodes, a smaller threshold takes longer, since it retains more strings.
As in the single-node benchmarks, larger operators exhibit better parallel efficiency, as fixed overheads become relatively less significant.
The hash optimizations of \appref{app:noderows} are also more effective at smaller node counts, reducing the baseline wall time and thereby lowering the measured strong-scaling efficiency.
These results show that $\F$-linear partitioning enables substantial parallel speedups within and across compute nodes.

\subsection{Two-dimensional transverse field Ising model}
\label{sec:tfim}

\begin{figure}[t]
    \centering
    \includegraphics[width=\linewidth]{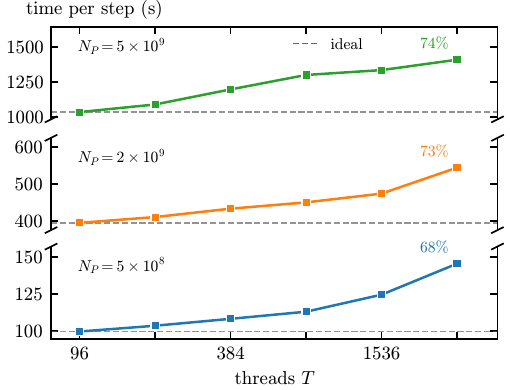}
    \caption{Weak scaling of the wall time per Trotter step of the $16\times16$ transverse-field Ising quench, at fixed per-node string budgets of $N_{P}=5\times10^8$, $2\times10^9$, and $5\times10^9$.
    Strings are distributed using the avoid $f = 50\%$ hash of \appref{app:noderows}, across AMD EPYC 9474F sockets, two of which live on each 96-core, $1.47\,\text{TiB}$ node. 
    At $32$ nodes, the largest configuration supports up to $1.6\times10^{11}$ Pauli strings, corresponding to $10.5\,\mathrm{TiB}$ of storage after deduplication and truncation, or $22\%$ of the available memory.}
    \label{fig:weakscaling}
\end{figure}

\begin{figure}[t]
    \centering
    \includegraphics[width=\linewidth]{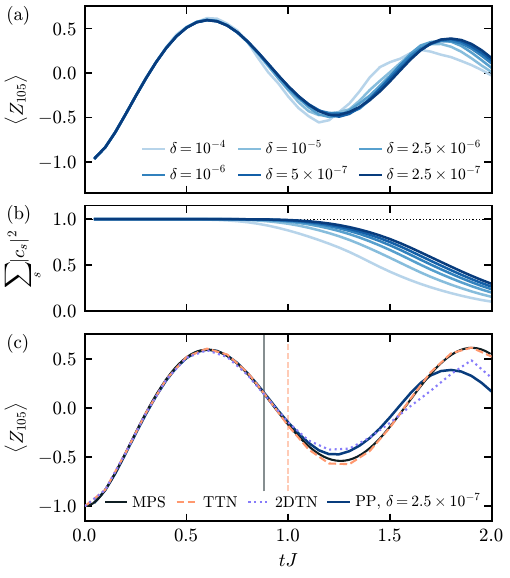}
    \caption{Central-site magnetization $\langle Z_{105}(t)\rangle$ for the $14\times14$ transverse-field Ising quench from $\ket{1}^{\otimes196}$ to $h/J=3$, using $40$ second-order Trotter steps with $J\delta t=0.05$.
    (a) Magnetization and (b) retained squared operator norm $\sum_s|c_s|^2$ for coefficient thresholds $\delta=10^{-4}$ to $5\times10^{-7}$.
    (c) Pauli propagation (PP) at $\delta=5\times10^{-7}$ compared with the MPS, TTN, and 2DTN-BP results of Ref.~\cite{vovrosh2026transverseising}.
    The Pauli propagation results are consistent up to the solid gray and dashed orange lines which indicate the symmetry-based convergence limits of MPS and TTN, respectively, from Ref.~\cite{vovrosh2026transverseising}.}
    \label{fig:quench}
\end{figure}

Next we consider quench operator dynamics in the two-dimensional transverse field Ising model on a square lattice with open boundaries, 
\begin{equation}
    H = -J\sum_{\langle jk\rangle} Z_j Z_k - h\sum_{j} X_j,
\end{equation}
where $J$ is the ferromagnetic coupling and $h$ is the strength of the transverse field.
Near the critical field $h_c/J\approx 3.04438$ \cite{Blote2002}, Pauli expansions of local operators grow rapidly \cite{begusic2025realtime}, making this system a demanding benchmark for evaluating weak scaling and the ability to exploit the combined memory capacity of multiple nodes. 
In \figref{fig:weakscaling}, we examine the weak scaling on a $16\times16$ lattice: starting from the product state $\ket{0}^{\otimes256}$ we quench to $h/J=3.04$ and compute the spatially averaged magnetization, $\langle Z(t)\rangle=256^{-1}\sum_j\langle Z_j(t)\rangle$, over $20$ second-order Trotter steps with $J\delta t=0.1$.
Instead of a coefficient threshold, we use the per-worker truncation scheme of \appref{app:parallel_ops}, fixing the retained-string budget per compute node $N_P$ and dividing it equally between two MPI ranks, one per NUMA domain.
Strings are distributed using the \emph{avoid} $f=50\%$ node hash described in \appref{app:hash}.
For weak scaling, we keep $N_P$, the string budget per thread, fixed as the number of nodes increases, with ideal scaling corresponding to constant wall time per Trotter step.
\figref{fig:weakscaling} shows the wall time per step from $1$ to $32$ nodes, with fixed per-node budgets of $5\times10^8$, $2\times 10^9$, and $5\times10^9$ strings.
At $32$ nodes, weak-scaling efficiency remains between $68\%$ and $74\%$, improving with the per-node string budget.
At the largest budget, $160$ billion strings, around $10$TiB  are retained after deduplication and truncation, with total memory usage peaking at approximately ${28.8\,\text{TiB}}$, out of the $47\,\text{TiB}$ available across $32$ nodes.
This demonstrates the ability to exploit the memory capacity of multiple nodes while maintaining efficient parallel scaling.

To assess how the increased memory capacity of our implementation improves convergence, we reproduce the $14\times14$ quench of Ref.~\cite{vovrosh2026transverseising}, starting from $\ket{1}^{\otimes196}$ and evolving the central-site magnetization $\langle Z_{105}(t)\rangle$ at $h/J=3$.
We evolve over $40$ second-order Trotter steps with $J\delta t=0.05$, with the only truncation being coefficient thresholds $\delta$ ranging from $10^{-4}$ to $2.5\times10^{-7}$.
As shown in Fig.~\ref{fig:quench}(a), decreasing the threshold extends the time range over which the magnetization converges.
At the smallest threshold, the expansion reaches $1.64\times10^{11}$ Pauli strings (about $11$TiB), distributed over $64$ nodes.
The corresponding retained operator norm, $\sum_s|c_s|^2$, is shown in Fig.~\ref{fig:quench}(b).
At $tJ=1$, the smallest threshold retains $99.1\%$ of the exact squared operator norm, indicating that only a small fraction of the operator weight has been discarded.
Finally, Fig.~\ref{fig:quench}(c) compares our Pauli propagation (PP) with results from Ref. \cite{vovrosh2026transverseising} using matrix product states (MPS), tree tensor network (TTN), and two-dimensional tensor networks with belief propagation (2DTN).
Within the convergence limits of MPS and TTN, indicated by vertical lines according to the symmetry-based criterion of that reference, all four methods agree.
At later times, the predictions diverge due to differing approximations, and the reference calculations no longer provide a converged benchmark for quantitative accuracy.
Together, these results show that distributed memory, usable with our implementation, enables substantially larger Pauli expansions while maintaining efficient parallel scaling, extending the range of accessible dynamics.

\section{Discussion and Outlook}
Pauli propagation is a competitive classical simulation method for quantum dynamics, but its scalability is typically constrained by the memory bandwidth and capacity required to manipulate large Pauli expansions.
In this work, we introduced an $\F$-linear partitioning scheme to enable scalability by distributing Pauli expansions across processors and compute nodes, such that gate applications are split into independent tasks with bounded communication.
This partitioning improves memory locality and enables the aggregate memory bandwidth and capacity of multiple nodes to be exploited, allowing larger simulations with efficient parallelization.

Both benefits of the partitioning concern memory.
Within a node, improved memory locality reduces data movement and enables more effective use of the available memory bandwidth.
Across nodes, the scheme enables the combined memory capacity to be exploited efficiently, requiring only point-to-point exchanges of Pauli strings rather than global redistribution.
Benchmarks on the kicked Ising circuit and the two-dimensional transverse field Ising model demonstrate substantial parallel speedups within and across compute nodes.
Our distributed-memory simulations reach Pauli expansions exceeding $10^{11}$ strings, extending the time range over which the simulated dynamics converge.

The remaining opportunities for performance improvements lie in reducing memory traffic and communication overhead.
For distributed execution, gates that couple workers still require pairwise exchanges and a brief synchronization, whose cost grows with the fraction of exchanging gates.
Tailoring the node hash to the circuit and number of nodes can reduce the fraction, although maintaining a balanced distribution of strings becomes increasingly challenging as the number of workers grows (see \ref{app:noderows}).
Single-node propagation remains limited by memory bandwidth, so further performance gains must come from hardware-specific optimizations that reduce the amount of data moved per string and gate, as discussed in \appref{app:engine}.

The partitioning strategy is independent of the underlying storage format and truncation scheme, allowing it to be incorporated into existing Pauli propagation implementations without altering their propagation algorithms.
A natural next step is to extend this approach to GPU-based implementations, where the benefit of the added memory bandwidth traditionally is hindered by the limited memory capacity.
Our methods thus provide the natural strategy for Pauli propagation methods to exploit modern parallel hardware, from the cores of a single node to many nodes.
With this, Pauli propagation makes for a competitive classical simulation method, enabling larger and more accurate simulations of quantum many-body dynamics.

\textit{Note added.}---During the preparation of this manuscript, related $\F$-linear partitionings were also introduced for the distribution over MPI ranks in \texttt{monoprop}~\cite{miller2026monoprop}.

\begin{acknowledgments}
L.D. thanks Nicolas Loizeau for the collaboration on PauliStrings.jl~\cite{loizeau2025quantum} that led to the idea for this project.
A.O. and A.K. acknowledge support from the Pierre Hohenberg Graduate Scholar Fellowship and computational resources provided by the Flatiron Institute. 
J.D. and L.D. are grateful for ongoing support by the Flatiron Institute.
The Flatiron Institute is a division of the Simons Foundation. 
\end{acknowledgments}

\clearpage
\appendix
\section{Linear hash details}
\label{app:hash}

\subsection{Linear partitioning}
\label{app:hashtheory}

A hash function $h\colon\F^{2L}\to\F^B$ assigns every Pauli string $P(\bv{p})$ to the partition $\partition{h(\bv{p})}$.
The hash is $\F$-linear if
\begin{equation}
    h(\bv{p}\xor\bv{q})=h(\bv{p})\xor h(\bv{q}) \quad \text{for all } \bv{p},\bv{q}\in\F^{2L},
    \label{eq:hashlinearity}
\end{equation}
Every such hash can be represented as $h(\bv{p})=M\bv{p}$ for a binary matrix $M\in\F^{B\times2L}$.
Each row $\bv{m}_j$ of $M$ selects a subset of bits from the binary representation $\bv{p}$, whose parity determines the corresponding partition label bit, $b_j=\bv{m}_j\cdot\bv{p}=\hw{\bv{m}_j\band\bv{p}}\bmod2$.
By linearity, the partition containing the identity string is the kernel $K=\ker M$, a linear subspace of dimension $2L-\rank M$. Every other partition is a coset of $K$, 
\begin{equation}
  \partition{\bv{b}}=\bv{p}_{\bv{b}}\xor K,~\text{with} ~ M\bv{p}_{\bv{b}}=\bv{b}.
\end{equation}
For full-rank $M$, all $2^B$ partitions therefore contain the same number $4^L/2^B$ of possible strings.
Moreover, multiplication by any Pauli string $Q = P(\bv{q})$ maps $\partition{\bv{b}}$ onto $\partition{\bv{b}\xor M\bv{q}}$ as a whole.

Conversely, this property characterizes linear partitioning up to relabeling.
For any $\bv{k}\in K$,  multiplication by $P(\bv{k})$ maps the identity to $P(\bv{k})$.
Since multiplication maps whole partitions onto partitions, it must map $K$ onto itself.
Thus $\bv{k}\xor K=K$ for every $\bv{k}\in K$, proving that $K$ is closed under $\xor$ and is therefore a linear subspace.
Similarly, multiplication by any $P(\bv{p})$ maps $K$ onto the partition containing $P(\bv{p})$, which is precisely the coset $\bv{p}\xor K$.
Since every linear subspace is the kernel of some linear map, these cosets define a $\F$-linear hash up to relabeling.

For common classes of gates, the offset space $D_U$ of \eqref{eq:offsetdimension}, which sets the number of partitions involved in each task, can be determined explicitly. 
A Pauli rotation maps every string to itself and to $QP$, so $W_U=\Span\{\bv{q}\}$ and $d_U\leq1$ regardless of the weight of $Q$.
When $M\bv{q}=\bv{0}$, no interpartition communication is required.
For a dense gate or channel acting on $k$ qubits, the offsets can span the full $2k$-dimensional binary space supported on those qubits, giving $d_U\leq\min(2k,B)$.

A Clifford gate maps each Pauli string to a single string up, $\bv{p}\mapsto S\bv{p}=\bv{p}\xor\Delta\bv{p}$ up to a sign, where $S=\Id+\Delta$ is a symplectic matrix.  
Therefore, propagation requires only updating the strings and routing them to their destination partitions, without generating additional strings or requiring deduplication.
The offset space is therefore $W_U=\im\Delta$, giving
\begin{equation}
    d_U=\rank(M\Delta)\leq\rank(S+\Id).
    \label{eq:cliffordbound}
\end{equation}
For the standard single- and two-qubit Clifford generators, $\rank(S+\Id)\leq1$ and $\rank(S+\Id)\leq2$, respectively, as listed in \tableref{tab:clifford_generators}.
For any $\bv{p}=\bv{p}_{\bv{b}}\xor\bv{k}$ with $\bv{k}\in K$, the output partition label is,
\begin{equation}
MS\bv{p}=\bv{b}\xor M\Delta\bv{p}_{\bv{b}}\xor M\Delta\bv{k}.
\end{equation}
Therefore, strings from the partition $\bv{b}$ can reach only partitions labeled by $\bv{b}\xor M\Delta\bv{p}_{\bv{b}}\xor M\Delta(K)$, at most $2^{d}$ of them with $d=\dim M\Delta(K)\leq d_U$. 
This routing depends only on the hash and the gate and can be precomputed.
For full-row-rank $M$, a right inverse $M^+$ determines the base offset $M\Delta M^+\bv{b}$ for each partition, while the columns of $M\Delta N$, where $N$ is a basis matrix for $K$, span the remaining possible offsets.

\begin{table}[t]
\caption{Bit updates of the standard Clifford generators on a string $\bv{p}=(\xp{p}\,\vert\,\zp{p})$, which leave all other bits unchanged, together with the bound $\rank(S+\Id)$ of \eqref{eq:cliffordbound} on the number of offset directions.}
\label{tab:clifford_generators}
\begin{ruledtabular}
\begin{tabular}{llc}
Gate & Bit update & $\rank(S+\Id)$ \\
\colrule
$H_j$ & $\xb{p}{j}\leftrightarrow\zb{p}{j}$ & $1$ \\
$S_j$ & $\zb{p}{j}\leftarrow\zb{p}{j}\xor\xb{p}{j}$ & $1$ \\
$\mathrm{CNOT}_{c\to t}$ & $\xb{p}{t}\leftarrow\xb{p}{t}\xor\xb{p}{c}$, \; $\zb{p}{c}\leftarrow\zb{p}{c}\xor\zb{p}{t}$ & $2$ \\
$\mathrm{CZ}_{jk}$ & $\zb{p}{j}\leftarrow\zb{p}{j}\xor\xb{p}{k}$, \; $\zb{p}{k}\leftarrow\zb{p}{k}\xor\xb{p}{j}$ & $2$ \\
\end{tabular}
\end{ruledtabular}
\end{table}

Linear hashes are also a classic universal family~\cite{carter1979universal}.
For a uniformly random $M$, two distinct strings collide with probability
\begin{equation}
    \Pr[M(\bv{p}\xor\bv{q})=\bv{0}]=2^{-B},
\end{equation}
since each row is independently orthogonal to the nonzero vector $\bv{p}\xor\bv{q}$ with probability $1/2$.
However, this collision guarantee holds only on average over $M$ and does not ensure balanced partitions for a fixed hash or fixed circuit.
For a fixed $M$ and circuit, partition balance depends on how the strings are distributed over the cosets of $\ker M$, as discussed in \appref{app:noderows}. 

\subsection{Implementation}
\label{app:hashimpl}
Evaluating the hash from scratch requires computing $B$ parities, costing $\mathcal{O}(BL)$ operations per string.
Once a partition label is known, linearity allows it to be updated incrementally, as $\bv{b}\mapsto\bv{b}\xor M\bv{d}$, where $\bv{d}$ is the offset generated by a gate.
For a gate acting on $k$ qubits, the $4^k$ possible offsets $M\bv{d}$ can be precomputed, reducing routing to a table lookup and XOR.

For distributed execution, the hash label is split into node and local components, $\bv{b}=(\bv{b}_{\mathrm{node}}\mid\bv{b}_{\mathrm{local}})$ corresponding to row blocks $M_{\mathrm{node}}$ and $M_{\mathrm{local}}$ of $M$.
The former assigns strings to workers, while the latter determines their partition within each worker.
The node part is fixed for the whole simulation, since changing it generally moves strings between all workers.
We draw the local rows at random, such that $M$ as a whole is a random full-rank matrix and every local partition can be used.

As the number of strings varies by orders of magnitude during a simulation, we dynamically adjust the number of local partitions to keep the mean partition size close to a target, as discussed in more detail in \appref{app:engine}.
The hash uses the first $B$ rows of a fixed, full-rank matrix with at most $2L$ rows, with the node-assignment rows placed first.
Increasing or decreasing $B$ activates or deactivates additional rows for local partitioning, leaving the node assignments unchanged.
Since each $B$-bit label is a prefix of the corresponding $(B+1)$-bit label, adding a row splits each partition $\partition{\bv{b}}$ into $\partition{(\bv{b},0)}$ and $\partition{(\bv{b},1)}$, requiring only one additional parity per string.
Removing the row merges these partitions back, without recomputing other label bits or transferring strings between workers.
This adaptation therefore requires no global synchronization.
Furthermore, each worker can adjust $B$ independently, coordinating only with workers coupled by a gate when strings must be exchanged.
At these exchanges, the participating workers adopt the largest of their $B$ values to ensure consistent partition labels.

\subsection{Choosing the node hash}
\label{app:noderows}
The choice of the node hash, $M_{\mathrm{node}}$, involves a trade-off between evenly distributing strings across workers and minimizing costly interworker communication.
We quantify the load imbalance by $\lambda=N_{\max}/\bar{N}$, where $N_{\max}$ is the maximum number of strings held by any worker and $\bar{N}$ is the mean number of strings per worker. 
For a Trotter step composed of Pauli rotations, we quantify communication by $f_{\mathrm{ex}}$, the fraction of gates satisfying $M_{\mathrm{node}}\bv{q}\neq\bv{0}$, which can exchange strings between workers. 
Unlike the load imbalance, $f_{\mathrm{ex}}$ depends only on the hash and the circuit, and is thus easy to compute beforehand.
We illustrate this trade-off using the $127$-qubit kicked Ising circuit, starting from the single-site observable $Z_{62}$.

The two objectives compete because the same gates that generate new Pauli strings also determine how they are distributed across workers.
Every string in the propagated operator is obtained from the initial observable $Z_{62}$ by multiplying with gate offsets $\bv{q}$, each shifting its node label by $M_{\mathrm{node}}\bv{q}$.
A gate with $M_{\mathrm{node}}\bv{q}=\bv{0}$ requires no interworker communication, but also cannot distribute strings to other workers.
If all gates are local, the node label is conserved, and all strings remain on the worker holding the initial observable, giving $f_{\mathrm{ex}}=0$ but $\lambda=P$.
Conversely, a uniformly random hash balances the strings, $\lambda\approx1$, but $f_{\mathrm{ex}}=1-1/P$, which tends to all gates as the number of workers $P=2^{B_{\mathrm{node}}}$ grows.

\begin{figure}[t]
\centering
\begin{minipage}[t]{0.49\columnwidth}\vspace{0pt}
\panellabel{a}{-5pt}{-8pt}{% Panel (a): Node-hash strategies on the 9-qubit heavy-hex lattice.
\begin{tikzpicture}[
    pp/picture,
    x=0.105cm,
    y=0.105cm,
    baseline=(current bounding box.north),
    pp/cell/.append style={line width=0.4pt}
]

    % ===== Row 1: y = -7.3 =====

    % Random
    \node[anchor=south, inner sep=1.5pt] at (9, -7.3) {random};
    \ppbits[pphash]{(0, -8.8)}{0,0,1,0,1,1,0,1,0,1,0,0,1,0,1,1,1,1}
    \ppbits[pphash]{(0, -9.8)}{1,1,0,1,1,0,0,0,0,1,1,0,0,1,0,0,0,0}
    \ppbits[pphash]{(0, -10.8)}{1,0,1,1,0,0,0,1,0,0,0,0,1,0,0,0,1,0}
    \draw[pp/cell] (9, -7.3) -- (9, -11.3);

    % X only
    \node[anchor=south, inner sep=1.5pt] at (30, -7.3) {$x$ only};
    \ppbits[pphash]{(21, -8.8)}{0,0,1,0,1,1,0,1,0,0,0,0,0,0,0,0,0,0}
    \ppbits[pphash]{(21, -9.8)}{1,1,0,1,1,0,0,0,0,0,0,0,0,0,0,0,0,0}
    \ppbits[pphash]{(21, -10.8)}{1,0,1,1,0,0,0,1,0,0,0,0,0,0,0,0,0,0}
    \draw[pp/cell] (30, -7.3) -- (30, -11.3);

    % ===== Row 2: y = -19.3 =====

    % Lattice cut
    \node[anchor=south, inner sep=1.5pt] at (9, -19.3) {lattice cut};
    \ppbits[pphash]{(0, -20.8)}{0,0,0,0,0,0,0,0,0,0,1,0,1,0,0,1,0,1}
    \ppbits[pphash]{(0, -21.8)}{0,0,0,0,0,0,0,0,0,0,0,1,1,0,0,0,1,1}
    \ppbits[pphash]{(0, -22.8)}{0,0,0,0,0,0,0,0,0,0,0,0,0,1,1,1,1,1}
    \draw[pp/cell] (9, -19.3) -- (9, -23.3);

    % Avoid f = 50%
    \node[anchor=south, inner sep=1.5pt] at (30, -19.3) {avoid $f = 50\%$};
    \ppbits[pphash]{(21, -20.8)}{0,0,1,0,0,1,1,0,1,1,0,0,0,0,0,0,0,0}
    \ppbits[pphash]{(21, -21.8)}{1,0,0,0,0,0,0,0,1,0,1,0,0,0,0,0,0,0}
    \ppbits[pphash]{(21, -22.8)}{0,0,1,0,0,0,1,0,1,1,0,0,0,0,0,0,0,0}
    \draw[pp/cell] (30, -19.3) -- (30, -23.3);

    % ===== Row 3: y = -31.3 =====

    % Sparse z, k = 2
    \node[anchor=south, inner sep=1.5pt] at (9, -31.3) {sparse $z$, $k = 2$};
    \ppbits[pphash]{(0, -32.8)}{0,0,0,0,0,0,0,0,0,0,0,0,1,1,0,0,0,0}
    \ppbits[pphash]{(0, -33.8)}{0,0,0,0,0,0,0,0,0,1,0,0,0,0,0,0,1,0}
    \ppbits[pphash]{(0, -34.8)}{0,0,0,0,0,0,0,0,0,0,0,0,0,0,0,1,0,1}
    \draw[pp/cell] (9, -31.3) -- (9, -35.3);

    % Z-parity (single row, vertically centered)
    \node[anchor=south, inner sep=1.5pt] at (30, -31.3) {$Z$-parity};
    \ppbits[pphash]{(21, -33.8)}{0,0,0,0,0,0,0,0,0,1,1,1,1,1,1,1,1,1}
    \draw[pp/cell] (30, -32.3) -- (30, -34.3);

\end{tikzpicture}}
\end{minipage}\hfill
\begin{minipage}[t]{0.49\columnwidth}\vspace{0pt}
\panellabel{b}{-2pt}{-8pt}{\includegraphics[width=\linewidth]{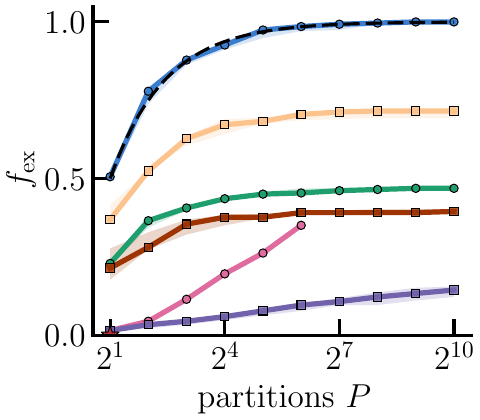}}
\end{minipage}\\[2pt]
\panellabel{c}{-5pt}{-5pt}{\includegraphics[width=0.49\columnwidth]{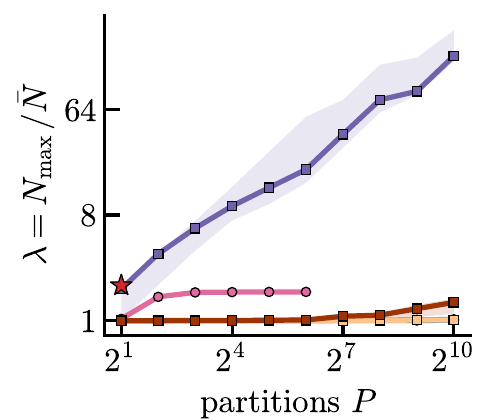}}\hfill
\panellabel{d}{-2pt}{-5pt}{\includegraphics[width=0.49\columnwidth]{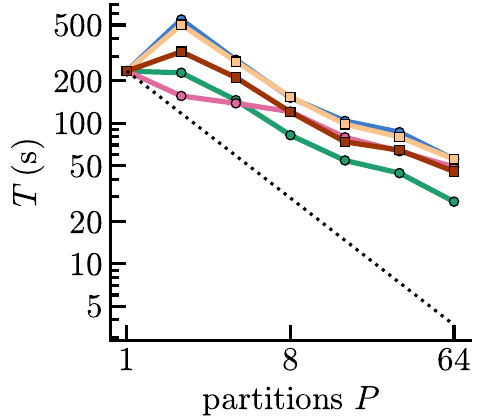}}\\
\includegraphics[width=\columnwidth]{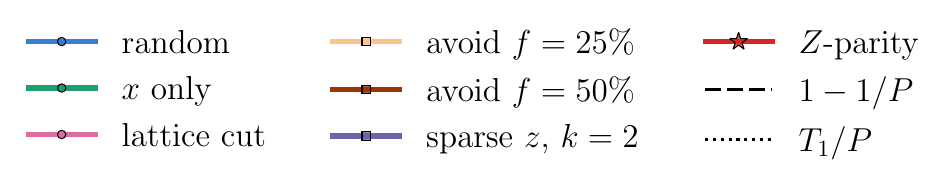}
\caption{Node-hash strategies for the kicked Ising circuit ($127$ qubits, $\theta_h=\pi/4$, $\delta=4\times10^{-6}$, $20$ Trotter steps).
(a) Example rows of $M_{\mathrm{node}}$ on a $9$-qubit heavy-hex lattice ($P=8$, except $Z$-parity with $P=2$).
Strategies: random (dense); $x$ only (dense on $x$ bits); lattice cut (the $z$ bits of contiguous lattice blocks, numbered in binary);
avoid $f$ (dense on coordinates outside the offsets of a random fraction $f$ of gates); sparse $z$ ($k$ randomly selected $z$ bits per row);
$Z$-parity (a single conserved parity row).
(b) Fraction $f_{\mathrm{ex}}$ of the $271$ gates of a Trotter step that can require interworker communication, versus the number of workers $P$.
(c) Load imbalance $\lambda=N_{\max}/\bar{N}$ versus $P$, obtained by distributing the same $3.6\times10^8$ strings from a single unpartitioned run after $20$ steps according to each hash.
(d) Wall time $T$ versus the number of workers $P$ (up to $8.2\times10^{8}$ strings), implemented as one MPI rank per NUMA domain of AMD EPYC 9474F nodes (two domains per node), using a single unpartitioned rank as the baseline.
Points in (b) and (c) are medians over three seeds for the random choices, with bands and bars spanning all three; points in (d) are the fastest of the three seeds.}
\label{fig:noderows}
\end{figure}
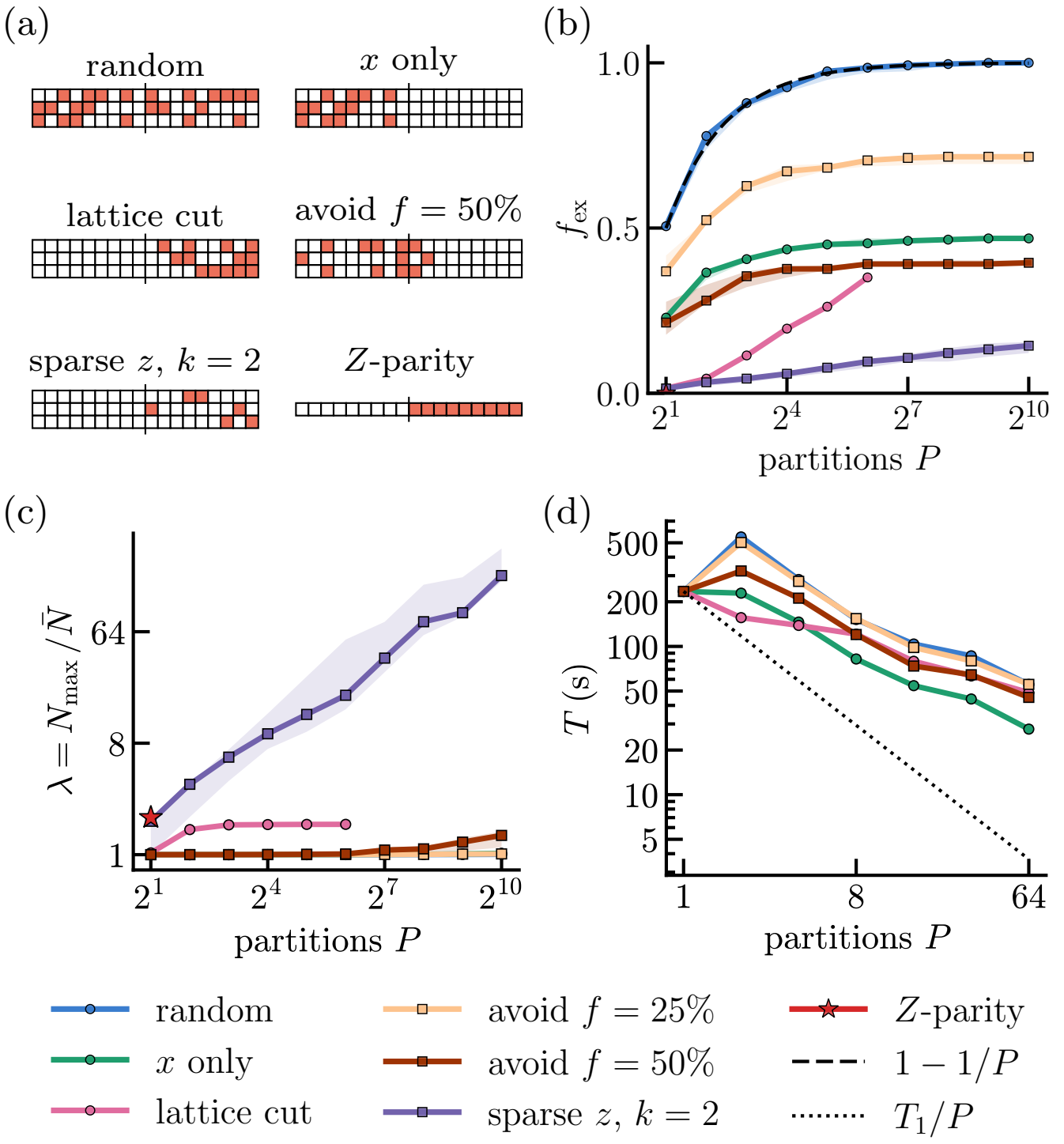

Between these extremes, the hash can be tailored to the circuit and available computational resources, for which we evaluate several families of options, as illustrated in \figref{fig:noderows}(a) for a small lattice.
For example, a random hash of only the $x$ bits makes all $ZZ$ rotations local, as their offsets only contain $z$ bits.
Alternatively, a \emph{lattice cut} divides the lattice into $P$ contiguous blocks, where row $i$ of the hash reads the $z$ bits of all qubits in blocks whose binary label has bit $i$ set.
Then, the $X$ rotations leave the $z$ bits unchanged, while $ZZ$ rotations within a block flip two $z$ bits whose contributions cancel.
Therefore, only $ZZ$ rotations connecting different blocks $\bv{c}$ and $\bv{c}'$ require communication, shifting the label by $\bv{c}\xor\bv{c}'$.
More generally, we can also define \emph{avoid} $f\%$, which selects a random fraction $f$ of the gates and construct a dense random hash using only coordinates outside their offsets, ensuring that these gates remain local.
Varying $f$ thus interpolates between an unrestricted random hash at $f=0$ and hashes that keep an increasing fraction of gates local, at the cost of fewer coordinates available for load balancing.

Tailoring the hash too much can lead to poor load balance, as illustrated by two common pitfalls.
The first arises from conserved quantities, illustrated by the $Z$-parity hash in \figref{fig:noderows}(a).
Every gate offset of the kicked Ising circuit has an even number of $z$ bits, so that the total $Z$-parity of a string, the parity of its number of $Z$ and $Y$ factors, is conserved.
A hash row reading all $z$ bits therefore avoids communication but assigns the same label bit to every propagated string, leaving one worker empty and giving $\lambda=2$ at $P=2$.
The same holds for any row, or combination of rows, that is orthogonal to every offset; such rows are attractive, as they never communicate, but must be avoided.

A second source of imbalance arises because propagated Pauli strings often have low weight, acting trivially on most qubits.
A hash row that reads only a few coordinates therefore evaluates to $0$ for most strings, concentrating them in a small number of partitions.
This is illustrated by the sparse $z$-only hash in \figref{fig:noderows}(a), where each row reads the $z$ bits of only $k$ qubits.
As $k$ decreases, fewer gates require communication, but the load imbalance increases, as seen in \figref{fig:noderows}(c) where at $P=64$ the $k=2$ hash gives $\lambda=19.6$, with $58\%$ of the workers empty.
The lattice cut, whose rows also read a restricted subset of coordinates, gives $\lambda=1.8$.
In contrast, the $x$-only hash and $f=50\%$ remain within $1\%$ of perfect balance, with $f_{\mathrm{ex}}=0.39$--$0.45$ compared with $0.99$ for a random hash [\figref{fig:noderows}(b)].
The same effect motivates drawing $M_{\mathrm{local}}$ densely at random to avoid nearly empty local partitions.

The reduced communication yields shorter runtimes, as shown in \figref{fig:noderows}(d).
On $8$ MPI ranks, a random hash is $1.6\times$ faster than a single rank, while the $x$-only hash, lattice cut and $f=50\%$ are $1.9$--$2.9\times$ faster, exchanging $1.0$--$2.9$\,TB instead of $5.9$\,TB.
On $64$ ranks, the random hash reaches $4.2\times$, against $4.8$--$8.5\times$ for these tailored hashes.

Tailoring the node hash is therefore particularly beneficial when the number of workers is small compared to the number of distinct gates in the circuit.
In this regime, the hash can keep a large fraction of the gates local while still distinguishing enough strings to maintain load balance, improving performance without affecting accuracy.
This is particularly relevant for a single-node job with only a few NUMA domains, such as the two domains in \secref{sec:kickedising}, where we use a lattice cut.
The hash families considered here illustrate how the gate and lattice structure can be exploited to reduce communication in other circuits, and if optimized to the problem and hardware used, can give large improvements.
As the number of workers grows, however, each additional row of $M_{\mathrm{node}}$ must distinguish strings not already separated by the previous rows while avoiding the offsets of gates kept local, making it increasingly difficult to maintain both low communication and good load balance.

\section{Truncation and measurement}
\label{app:parallel_ops}
In \secref{sec:methods}, we showed how using a linear hash avoids global communication by decomposing gate applications into independent tasks involving small groups of partitions. 
However, both measurements and truncation can reintroduce global communication.
Here, we show that common truncation schemes and measurements
can be performed locally on each worker, with communication
limited to scalar or fixed-size histogram reductions.

Various truncation schemes can be deployed to control the growth of the Pauli expansion, discarding strings expected to contribute little and thus trading accuracy for memory. 
For example, one can truncate with respect to coefficient weight after every operation, either removing all strings with coefficients under some threshold $\delta$, or truncating down to only the $N_{\max}$ largest coefficients~\cite{begusic2024fast,begusic2025realtime,loizeau2025quantum,rudolph2025pauli}. 
Alternatively, one can keep only strings with spatial length below a threshold, add depolarizing noise channels to exponentially suppress long strings~\cite{schuster2025polynomial}, or replace the expansion at certain times with an unbiased resampling of existing strings down to $N_{\max}$~\cite{barron2026observableestimationabsenceclassical,rudolph2026pushingmemorybarrierpropagation}.
In all cases, truncation only removes strings, leaving the partitions of the retained ones unchanged.
Schemes that evaluate criteria independently for each string, such as coefficient, weight, or length thresholds, are therefore trivially parallel with each partition applying them locally. 

\begin{figure}[t]
\includegraphics[width=\linewidth]{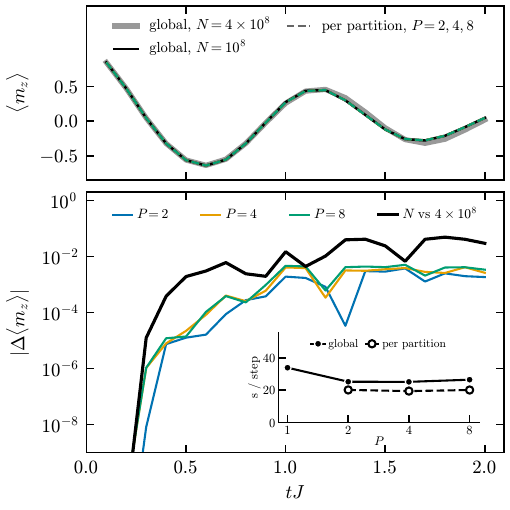}
\caption{Retaining the globally largest $N$ coefficients against retaining the $N/P$ largest on each of $P$ workers, for the $16\times 16$ transverse-field Ising quench of \secref{sec:tfim}: (a) the magnetization for both truncations at each $P$, (b) their difference next to the effect of the truncation itself (a $4\times$ larger budget), and (c) the wall time per Trotter step.}
\label{fig:truncation_equivalence}
\end{figure}

Keeping the $N_{\max}$ largest coefficients is useful when the total computational resources available is fixed, but it is not trivially parallel: it requires identifying a global cutoff, and therefore global communication over all the data.
We introduce two approximations that progressively reduce this cost and that apply equally to keeping the $N_{\max}$ shortest strings.
In the first method, each worker bins its coefficient magnitudes logarithmically, and the histograms are summed in a single reduction whose size is independent of the number of strings.
All strings in bins above the cutoff are retained, while the remaining budget is filled from the cutoff bin.
Only the strings retained from this boundary bin then depend on the hash.
Alternatively, each of the $P$ workers can apply the same rule to a local budget $N_P = N_{\max}/P$, eliminating communication and bounding the memory used by each worker.
This retains the largest coefficients on each worker rather than the largest overall, so the result depends on the distribution of strings across workers and therefore on $P$.
This second approach was employed in \figref{fig:weakscaling}, and \figref{fig:truncation_equivalence} compares the two approaches at the same total budget, showing they give equivalent magnetization in less time per trotter step in the transverse field Ising quench.
The difference between them remains well below the effect of the truncation itself, estimated from a $4\times$ larger budget [\figref{fig:truncation_equivalence}(b)], indicating that per-worker truncation closely approximates global truncation for this benchmark.

Now we discuss how to apply parallelization to measurements. The expectation value of a propagated observable in the state $\rho$ is
\begin{equation}
    \langle O(t) \rangle = \operatorname{Tr}\left(\rho\, O(t)\right) = \sum_i c_i \operatorname{Tr}\left(\rho\, P(\bv{p}_i)\right).
\end{equation}
Individual string contributions can be evaluated efficiently from their symplectic representation for many initial states of interest. 
For computational basis states, only Pauli strings containing $I$ and $Z$ contribute, with expectation values determined by a bitwise parity. 
For general product states, expectation values factorize into single-qubit contributions, costing $\mathcal{O}(L)$ per string. 
For stabilizer states, contributions are $0$ or $\pm1$ and can be evaluated through stabilizer-group membership tests using binary linear algebra. 

Each partition independently accumulates its contribution to the expectation value, and a single scalar reduction then sums these contributions.
Operator overlaps can be evaluated similarly, since only strings present in both operators contribute to the Hilbert--Schmidt inner product.
When both operators use the same node hash, identical strings reside on the same worker, such that their intersection is computed locally. 

\section{Implementation details}
\label{app:engine}
The partitioning of \secref{sec:methods} determines which strings interact and which workers communicate, but leaves open how each task is executed.
Since operations on individual Pauli strings are inexpensive, the performance of \texttt{paulistrings-rs} is primarily determined by memory traffic.
We reduce this traffic through compact storage, cache-sized working sets, selective processing of strings, gate fusion, and bookkeeping integrated into existing passes.
The resulting propagation loop of a single worker is summarized in Algorithm~\ref{alg:engine}.

\begin{algorithm}[t]
\caption{Propagation loop for a single worker.
A run consists of consecutive gates that exchange no strings, or a single gate that can exchange strings between workers.
The number of active local hash bits, $B\leq B_{\mathrm{local}}$, is adjusted dynamically to maintain a target partition size, with $2^B$ local partitions per worker.}
\label{alg:engine}
\footnotesize
\algrenewcommand\algorithmicindent{1.2em}
\hrule\vspace{3pt}
\begin{algorithmic}[1]
\For{each run $R$ of consecutive gates}
    \Statex \hspace{\algorithmicindent}\textit{Prepare}
    \If{$R$ exchanges strings}
        \State agree on $B$ with the workers that $R$ couples
        \State send strings marked for exchange in chunks
    \Else
        \State split or merge partitions towards the target size
    \EndIf
    \Statex \hspace{\algorithmicindent}\textit{Apply, in parallel by work stealing}
    \For{each group of partitions that $R$ couples}
        \State load the group once its incoming strings have arrived
        \State apply each gate, skipping unchanged partitions
        \State accumulate the histogram and mark strings for the next exchange
    \EndFor
    \Statex \hspace{\algorithmicindent}\textit{Truncate}
    \If{the policy keeps the globally largest coefficients}
        \State reduce the histograms, decide the cut for the next Apply
    \EndIf
\EndFor
\Statex \textit{Finish}
\State agree on $B$ across all workers and apply the pending cutoff
\end{algorithmic}
\vspace{3pt}\hrule
\end{algorithm}

Each Pauli string is packed into $2\lceil L/64\rceil$ 64-bit words for efficient bitwise operations.
Within each partition, the $x$ bits, the $z$ bits, and the coefficients are stored in separate arrays, with entries sorted by Pauli string.
Threshold truncation therefore requires only a single pass and preserves the ordering of the retained strings.
Under the phase convention of \secref{sec:methods}, every encoded Pauli string is Hermitian, so Hermitian observables have real coefficients.
These remain real under Hermiticity-preserving gates, allowing coefficients to be stored as real rather than complex numbers, halving their memory footprint.

\begin{figure}[t]
\includegraphics[width=\linewidth]{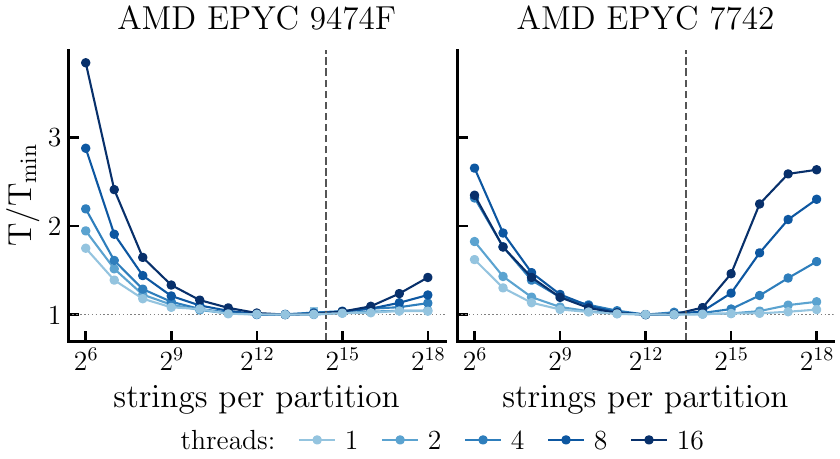}
\caption{Wall time of the kicked Ising circuit ($127$ qubits, $20$ Trotter steps, $8.3\times10^6$ strings) against the target number of strings per partition.
Results are shown for $1$ to $16$ threads within one NUMA domain of AMD EPYC 9474F (left) and 7742 (right) processors.
Wall time $T$, normalized by the minimum $T_{\min}$ over target partition sizes for each processor and thread count.
Dashed lines indicate the L2 cache capacity per core, assuming $48$ bytes per string.
Performance is near-optimal for a broad range of partition sizes, with little dependence on thread count.}
\label{fig:buckets}
\end{figure}

Partitioning also reduces memory traffic for a single worker by shrinking the working set for deduplication from $N$ to roughly $2^{d_U} N / 2^B$ strings.
With enough partitions, this working set fits in the processor caches, so hash-map accesses or sorting operations avoid much of the traffic to main memory.
The number of local partitions is therefore adjusted dynamically to keep the working sets small without creating excessive partitioning overhead.
An additional local hash row is activated when the mean partition size exceeds a target, splitting every partition as described in \appref{app:hashimpl}, and deactivated only when the mean falls below one eighth of the target, preventing frequent splitting and merging.
\figref{fig:buckets} shows the wall time as a function of the target partition size on two nodes with different L2 cache sizes.
Small partitions incur additional overhead, while large partitions reduce cache locality, resulting in a broad performance optimum that changes little with thread count.
By default, we target $2^{10}$ strings per partition and maintain at least $2^7$ local partitions to ensure parallelism for small operators, independently of the number of threads.

We also avoid unnecessary memory accessing by exploiting that Pauli rotations affect only strings anticommuting with their generator $Q$ [\eqref{eq:paulirotation}].
Each partition maintains bitwise OR summaries of its $x$ and $z$ bits, allowing partitions that commute entirely with $Q$ to be skipped without reading their contents.
Within affected partitions, one pass identifies the anticommuting strings, while a second updates their coefficients and generates partners $QP$, which belong to the partition $\bv{b}\xor M\bv{q}$.

The generated partners must then be combined with existing strings.
XOR with the fixed offset $\bv{q}$ reverses the order of two strings exactly when $\bv{q}$ sets the first bit in which they differ.
The partners of a sorted partition can therefore be arranged in sorted order by block swaps, without a comparison sort.
Since the partner $QP$ also anticommutes with $Q$, it can coincide only with an existing anticommuting string in the destination partition, already identified in the first pass.
These contributions are summed directly, new strings below the coefficient threshold are discarded, and the remainder is merged into the sorted destination, copying unchanged strings in contiguous blocks.
For a Clifford rotation, with $\theta$ an odd multiple of $\pi/2$, every anticommuting string is instead replaced by its partner, and no coefficients have to be summed.

Since every pass over the strings costs memory bandwidth, consecutive gates that exchange no strings between workers are applied in a single pass.
Each gate $U$ couples only partitions within a coset of its offset space $D_U$, so a sequence of gates couples only partitions within a coset of the sum of their offset spaces.
Each such group is loaded into cache, propagated through the entire sequence with deduplication after every gate, and written back.
Adding gates can double the number of coupled partitions, so fusion stops when a group would exceed eight partitions, keeping the working set small enough to benefit from the L2 cache.
Longer sequences can instead be fused by delaying deduplication, propagating strings along branching Pauli paths before merging identical strings.
This trades fewer memory passes for larger intermediate expansions, moving from breadth-first towards depth-first propagation.
For runs of Clifford gates, such as the $ZZ$ rotations of the kicked Ising circuit, each string is propagated through the entire run, assigned to its final partition, and sorted only once.

Since fusion eliminates intermediate passes over the strings, truncation schemes requiring global information between gates would generally retain different strings if applied only at the end of a run.
Our implementation therefore fuses only under the trivially parallel criteria of \appref{app:parallel_ops}, for which the result is unchanged.

For truncation schemes requiring global coordination, we instead reduce memory traffic by integrating bookkeeping into existing passes.
When approximating the globally largest coefficients, each task accumulates a histogram of squared coefficient magnitudes in power-of-two bins while its output partitions remain in cache.
After the histograms are reduced, the resulting cutoff is applied when the next gate loads each partition, avoiding a separate truncation pass.
Communication is prepared similarly.
If the next gate exchanges strings, the current pass marks those that must be sent, so the subsequent exchange reads only the marked strings.
These are sent in chunks using non-blocking point-to-point messages, allowing each worker to process a group as soon as its incoming data arrives, overlapping communication with computation.

\bibliography{bibliography}

\end{document}